\documentclass[11pt]{article}

\usepackage{psfrag,graphicx}
\usepackage{amssymb}
\usepackage{amsfonts}
 \newcommand{\be}[0]{\begin{equation}}
   \newcommand{\ee}[0]{\end{equation}}
   \newcommand{\ba}[0]{\begin{eqnarray}}
   \newcommand{\ea}[0]{\end{eqnarray}}

\begin{document}

\setlength{\baselineskip}{18pt} \thispagestyle{empty}
\title{Target mass correction, based on the re-scaled parton densities}
\author{\textbf{M.~M.~Yazdanpannah$^{(1)}$\footnote{e-mail:\texttt{myazdan@uk.ac.ir}}, R.~Mohammadi pour$^{(1)}$ \footnote{e-mail:\texttt{reyhane8412@yahoo.com}} and A.~Mirjalili$^{(2)}$
\footnote{e-mail:\texttt{\texttt{a.mirjalili@yazd.ac.ir}}}}}
\date{\vskip10mm\emph{$^{(1)}$
Faculty of physics, Shahid Bahonar University of Kerman, Kerman,
Iran\\$^{(2)}$Physics Department, Yazd University, P.O.
Box 89195-741,  Yazd, Iran }}
\maketitle \thispagestyle{empty}
\begin{abstract}Target mass correction (TMC) is being used  to improve the theoretical results for
the nucleon structure functions. This improvement makes the
fitting  more reliable with the available experimental data
in lepton scattering off the nucleon . The recent relations, using  the Georgi and  Politzer approach  indicate
that employing   TMC effect are cumbersome, long and contains
complicated integrals. Some of these integrals can not be solved
exactly and people are forced to use approximate methods which
seems  not to give sufficient and  precise results. On the other hand on analysing
the TMC effect we encounter with the threshold problem for nucleon structure
which results to have the structure function in a regions which are physically forbidden.
Here we render  a new solution way, based on the  re-scaled parton densities to resolve the difficultly,
relating to the TMC effect. Our solution way can be employed
directly and easily to the parton distribution rather than to the
nucleon structure function. There are different options to amend
the parton densities. What we call as third re-scaling is the best
one. Using it we are able to discard the threshold problem for the
nucleon structure function without resorting to apply some auxiliary mathematical tools like the step
function which is plausible by people .  On this way we obtain a better agreement for the nucleon
structure function  with the available experimental data in
comparison with the ordinary TMC results, based on Georgi and Politzer
approach and some other approaches.
\end{abstract}

\newpage
\renewcommand{\thepage}{\arabic{page}}
\pagestyle{plain} \setlength{\baselineskip}{16pt}
\setlength{\parskip}{11pt} \setcounter{page}{2}


\section{Introduction}
For the very small size of dimension, smaller scale than nuclei
size, there are some specific theories  to investigate the parton
interaction dynamic \cite{gel,feyn,bro}. However there is not any
completed and exclusive theory to describe exactly what are being
happened inside the nucleon. To recognize  the subatomic
structure better we need to resort the deep inelastic scattering (DIS) process
for leptons off the nucleon as a sufficient and  rich investigating instrument. The DIS
experiment can be used to investigate the dynamical behavior which
are govern the quarks and gluons interactions inside the nucleon.

To analysis the experimental data of DIS, there are different
models. But these models are not exact and need some corrections
to yield us reliable results. In the related  approximations  the
forces between the partons which are infinitesimal and do not leak out  at any place,
are usually ignored. If we wish to achieve more exact relations
for describing the subatomic particles, we need to consider the
effect of these forces in the interval which are more effective.
As a result we should employ the TMC effect which can be applied
in the naive quark-parton model.

TMC is in fact to consider the effect of mass for the particles in DIS
experiment which people usually ignore it. The history of
considering TMC is too long and back essentially to 1976 when
Georgi and Politzer  took  it into account to the nucleon
structure function which is relating as well the electro-weak
force \cite{2}. TMC in DIS scale, includes the effects resulted
from massive quarks. This correction was followed latter on by
Elice, Formansky and Petronzio, based on quark-parton model
\cite{efp}. All the calculations have been done in the leading
order(LO) approximation which  then extended to the
next-to-leading-order(NLO) approximation by De Rujula, Georgi and
Politzer \cite{5}. Recently some investigations have been done by
Kretzer and Reno in the NLO approximation , considering the
charged current which exists when  the partons are assumed to
interact with each other \cite{kr,1}.

The  method which was introduced by Georgi and Politzer, was in
fact based on operator product expansion (OPE) theorem \cite{4}.
Latter on Nachtmann \cite{n} showed that using OPE theorem in
terms of $ \frac{1}{Q^{2}} $ expansion, only a twist two operators
are needed where  twist is defined as dimension minus spin.  The dimension means
the naive canonical dimension of an operator and spin relates  to its transformation properties
under the Lorentz group. In general, power-suppressed contributions are referred  as
higher-twist contributions. TMC consists of power-suppressed twist-2 operators that
occurs in the expansion when the target mass is not set to zero. Those
corrections can be exactly taken into account by  expressing
the structure functions in terms of Nachtmann
variable, $\xi$, rather than  the x-Bjorken variable. At a specific $Q^{2}$, this variable is defined
by a four vector which is  relating to the momentum fraction  of
nucleon (p), carried by parton(k)\cite{4}:
\begin{equation}\label{1}
  \xi(x,Q^{2})
  =\frac{k^{0}+k^{z}}{p^{0}+p^{z}}=\frac{2x}{1+\sqrt{1+\frac{4x^{2}M^{2}}{Q^{2}}}}\;.
\end{equation}

Using this variable, there are some difficulties whose famous one
is called ``threshold problem". When nucleon structure function is
written in terms of Nachtmann variable, the maximum value of $ \xi
$ at a specified value of $ Q^{2} $ is $ \xi_{0}=\xi(x=1)< 1 $
which causes the parton distribution and finally nucleon structure
function appear  on the physical range of $ \xi=0 $ to $
\xi=\xi_{0} $. For the interval in which $ \xi_{0} < \xi < 1 $,
the Bjorken variable $ x $ exceeds its maximum value $ x=1 $ which
is physically forbidden. To overcome this difficulty, some
solution ways are suggested which by themselves makes other
difficulties \cite{4,7,6}. On the other hand if someone  wishes to follow the Georgi and Politzer approach then  the main relations to
consider the  TMC effect  for the structure function involve many complex integrals
which we are not able  to calculate  them easily.

Here we first introduce the recent relations which exist for TMC
and then express our idea to resolve the difficulty which we
referred it. Finally we present the result of our method  which
contains the TMC effects. To describe briefly, our suggested way
is to make some changes on parton distributions, based on the
re-scaling the variable of momentum fraction which are carried
by partons. This re-scaling will finally affect the nucleon
structure function to indicate a proper feature

The organization of this paper is as follows. In Sec. 2 we
consider the TMC effect based on  the Georgi and Politzer approach. We discuss in Sec. 3 the reasons and motivations  which lead
people to take into account  the approaches, based on the modified parton densities to  employ  the
TMC effect. In Sec. 4 we present our suggested way for the
re-scaled parton densities to give us the reliable results. Some
discussions and  comparisons between our results with the results
of the other methods and experimental data are done in Sec. 5. Finally
we give our conclusion in Sec. 6.

\section{Revisiting the TMC; Georgi and Politzer  approach}
Here we intend to consider where the TMC has been ignored in which
we need to amend the nucleon structure function to involve the
mass corrections. In fact we are looking for where the required
forces in interaction between partons  have been existed in which we need to employ the
corrections. If we use Lorentz frame in which $
\vert\overrightarrow{p\vert} \gg m,M $, i.e. the frame where
parton has infinite momentum then we can take the nucleon and its
parton as massless particles. In this frame, the relativistic
delay time, possesses low rate for parton interactions.  Therefore virtual photon interacts
in a short time with quarks which seems like free particles. So in a naive  theoretical consideration of  DIS
processes, quarks do not interact with each other. In a real case when partons are constrained
to be inside the nucleon, they interact to each other. Effect of the forces which quarks make to each other,  would causes to
produce some small mass effect which is related to TMC.
The relations of the nucleon structure
function should be  amended, considering the required mass effect of target.

According to Ref.\cite{4}, the TMC effect for nucleon structure
functions can be related to the limit of these functions when we do
not consider any mass effect. The related equation is called
``master equation". This equation  was
first obtained by Georgi and Politzer (GP) \cite{2}. This equation is
written for the unpolarized case at twist 2 order by:
 \begin{eqnarray}\label{2}
 F^{TMC}_{2}(x,Q^{2})=\frac{x^{2}}{\xi^{2}r^{3}}F^{0}_{2}(\xi,Q^{2})
 \nonumber\\+\frac{6M^{2}x^{3}}{Q^{2}r^{4}}h_{2}(\xi,Q^{2})+\frac{12M^{4}x^{4}}{Q^{4}r^{5}}g_{2}(\xi,Q^{2})\;.
\end{eqnarray}
The related expressions in above equation are as following:
\begin{equation}\label{3}
  h_{2}(\xi,Q^{2})=\int^{1}_{\xi}du
  \frac{2F_{2}^{0}(u,Q^{2})}{u^{2}}\;,
\end{equation}
\begin{eqnarray}\label{4}
 g_{2}(\xi,Q^{2})=\int^{1}_{\xi}du h_{2}(u,Q^{2})
 \nonumber\\=\int^{1}_{\xi}dv
 (v-\xi)\frac{F_{2}^{0}(v,Q^{2})}{v^{2}}\;.
\end{eqnarray}
The `$r$' variable  in these equations is given by:
\begin{equation}\label{5}
 r=\sqrt{1+\frac{4x^{2}M^{2}}{Q^{2}}}\;.
\end{equation}
We should note that the coefficients which exist in Eq.(\ref{2})
like $ \frac{x^{2}}{\xi^{2}r^{3}} $ , $
\frac{6M^{2}x^{3}}{Q^{2}r^{4}} $ and $
\frac{12M^{4}x^{4}}{Q^{4}r^{5}}$ are identical  at
all LO, NLO and higher approximations. As can be seen the master
equation, Eq.(\ref{2}), is a long equation and involves
complicated integrals, given by Eqs.(\ref{3},\ref{4}). So
employing the  TMC effect  for nucleon structure function by this
approach, is a tedious task. Sometime due to the unsolvable
integrals, achieving the correction, using this approach is
impossible. On this base, some people are trying to use the
approximate relations \cite{5}. But these approximations would
yield us reliable results just at some specific energy scales.
This is why people are encouraging to seek the other approaches to employ
the TMC effect as we discuss them in the next section.
\section{Parton densities and TMC, motivations}
Here we intend to express the motivations which make the people to consider
the TMC effect through the parton densities. Before that we need to some basic definitions.

The two standard moments of structure functions which are used  in the
literature are the Cornwall-Norton (CN) and Nachtmann moments
\cite{cn}. The CN moments of $F_2$ are given by:
\begin{eqnarray}
M_2^n(Q^2) &=& \int_0^1 dx\ x^{n-2}\ F_2(x,Q^2)\ , \label{e2}
\end{eqnarray}
which are appropriate for the region $Q^2 \gg M^2$.

On the other hand, the Nachtmann moment contains  $M^2/Q^2$
corrections to the Bjorken limit, and are given by:
\begin{eqnarray}
\mu_2^n(Q^2) &=& \int_0^1 dx\ { \xi^{n+1} \over x^3 } \left[ { 3 +
3(n+1)r + n(n+2)r^2 \over (n+2)(n+3) } \right]
\nonumber\\
&&F_2(x,Q^2)\ ,
\label{e9}
\end{eqnarray}
where $r$ is defined by Eq.(\ref{5}) and $\xi$ is representing as
before the Nachtmann variable. A particular feature of these
Nachtmann moments is that they are supposed to factor out the
target mass dependence of the structure functions in a way such
that its moments would equal the moments of the corresponding
parton distributions.

Recent experimental data includes the region of large Bjorken $x$.
Therefore we need to achieve a better control to employ parton
distributions in this region. This region contains higher twist
contributions and also the corrections  which are relating to the
mass of target. So we should be careful in  the meaning of  parton
distributions for such case.

To extract parton distributions from the measured structure
functions we need  a prescript to include the TMC effect and in this regard there are
different approaches. One of these approaches  as was referred before, is the approach of  Georgi
and Politzer  which is based on the operator product expansion theorem. In this case in  the related expansion we keep  the terms which are  proportional to ratio of $\frac{m^2}{Q^2}$ where $m$ is refereeing to  the mass of
target. The difficulty which exists in this approach is that the parton densities are used in a region which is physically forbidden. If we consider the mass of target in  calculations then the maximum value for the fraction of momentum is not one. In fact the parton densities are not defined when the  momentum fraction goes to 1. The maximum value for this fraction is given by:

\begin{equation}\label{6}
\xi_0=\xi (x=1)=\frac{2}{1+\sqrt{1+\frac{4 M^{2}}{Q^{2}}}}\;.
\end{equation}
which is smaller than 1 for any finite $Q^2$. It seems that if TMC
are included, one loses the partonic interpretation: no parton
distribution with TMC can be really defined. Given the importance
of the potential consequences, it is urged that more investigation
on this fascinating subject be made.

\section{TMC, based on  the re-scaled partons}
As was discussed in above, however GP approach give us reliable
results for nucleon structure function (SF) but when people are
intending to re-express this effect in this approach in terms of
parton densities, they encounter with parton densities in an
un-physical region. Additionally  this approach by itself contains
cumbersome and unsolvable analytical  integrals which forces  people to use
the approximation methods to consider this approach. In order to discard this deficiencies, it is better
to do directly the mass correction on parton distributions rather than to use
Eq.(\ref{2}). In Eq.(\ref{2}) the TMC is done only on the nucleon
SFs, and  the mass correction has not been done by  parton
densities.

The main question is how the mass correction can be done on parton
distributions in which the resulted correction is equal to what
have been obtained, using  Eq.(\ref{2}) such that there is not the
threshold problem for the nucleon structure functions. Therefore
we are going to employ the mass correction on parton distributions in
which the nucleon structure function, $ F_{2} $, resulted from
this correction, is very similar to what is obtained from
Eq.(\ref{2}). The suggested solution  is that to do the mass correction for
parton distributions, using a  method which is called
re-scaling. We mean from re-scaling to substitute new variable
instead of the variable which is usually  used to describe the
parton distributions. In fact we replace the Bjorken variable-$x$
with the new one in parton distributions and investigate the
changes which are resulted.

Now we use the parametrization of Ref.\cite{10} for the
parton distributions, and then employing the re-scaling
prescription. We could  get finally the mass correction for the
nucleon structure function. But what is the proper re-scaling and
what variable
should be replaced instead of $x$-Bjorken variable?.\\
There are three types of re-scaling:\\
I) Replacement the $x$ Bjorken variable with the Nachtmann
variable, $\xi$, in the whole part of the parton distributions. If
we assume the following form for  the parton densities:
\begin{equation} q(x,Q^2)=ax^b(1-x)^c(1-d x^{0.5}+e x) \label{p1} \end{equation} then
the re-scaled presentation for parton densities in this type of
re-scaling,  is appeared as
\begin{eqnarray} a\xi^b\left(1-\xi\right)^c(1-d \xi^{0.5}+e \xi) \label{p2}
\end{eqnarray}
 \\
II) Replacement the $x$ Bjorken variable with the Nachtmann
variable, $\xi$, in the whole part of the  parton distributions
but substituting the maximum value of transferred momentum
fraction , 1, by $\xi_0(x=1)$. So the re-scaled presentation of
the second type for parton densities will be as
\begin{eqnarray} a\xi^b\left(\xi_0(x=1)-\xi\right)^c(1-d \xi^{0.5}+e \xi) \label{p3}
\end{eqnarray}
\\
III) Replacement the $x$ Bjorken variable with the Nachtmann
variable, $\xi$, in all parts of the  parton distributions except
for the (1-$x$) part.

We believe that the third type of  re-scaling is the best one as the
results which are arising  out from this re-scaling confirm this
reality. In this case the threshold problem would be solved
automatically and the nucleon structure function goes to zero when
the $x$ is approaching  to 1.  By keeping the  $(1-x)$ term
without any change in parton distribution function, in fact we do not allow
the structure function achieves numerical values  when $x$-Bjorken
variable is grater than 1. It is because that the unchanged $(1-x)$
term implies  this fact that maximum value of $x$ is 1. Our
justification for this re-scaling is that when the transferred
 momentum fraction which is carried by partons achieves its maximum
value, 1, only the three valence quarks exist and there
is no chance for sea quarks and gluons to be appeared. On the
other hand on this limit, there is not any partonic interactions
to produce sea quarks and gluons. If such interactions do not
exist it means that there is not any mass effect. In this case we
can assume that $\xi$ and $x$ variables are equal to each other. This is why we
keep the  (1-$x$) term in this re-scaling without any change.

For more clarifying the subject, we can say
however TMC is  an overall  kinematic problem but this  does  not mean that the quark-quark (qq) correlator
and also  quark-quark-gluon  (qqG) and qqGG correlators do not contribute in the TMC effect. As has been written
in \cite{DAM}, due to  off-shellness of the quark in a direct consequence of QCD we do not need to
consider the qqG and qqGG correlators but in on-shellnes case we need to take these correlators as
Ellis, Furmanski and Petronzio  used them in their article  \cite{EFP}.
Accepting the role of the correlators in TMC effect,  means that we should consider the parton
interactions for this  effect.

In summary, if we consider the expression $ax^b(1-x)^c(1-d x^{0.5}+e x)$ as
the parameterized form for parton distribution then the re-scaled
form would be as:

\begin{eqnarray}a\xi^b(1-x)^c(1-d\xi^{0.5}+e\xi)\;.\end{eqnarray}

In this expression the first term is controlling the behavior of parton
distribution at low $x$ values. The second term is relating to
behavior of parton distribution at large values of $x$ which we
keep it not to change. The last term is controlling the behavior of
parton distribution at medium values of $x$ variable which
changes to $\xi$  during the re-scaling processes. When this type of
re-scaling is used, we do not need to use the step function to
discard the threshold problem as has been  used in Ref.\cite{4}. On
the other hand,  the unchanged $(1-x)$ term which is used in the
third type of re-scaling, will guarantee to yield  us a proper result for
the structure function with acceptable threshold behavior.

After employing the proper re-scaling on the parton densities, it
is now turn to construct the nucleon structure function, $ F_{2} $, which
at LO and NLO approximations  are given respectively by \cite{10}:
\begin{equation}\label{6}
 F^{ep}_{2,LO}(x)=\sum_{i}(e_{i}^{2}xf_{i}(\xi))
\end{equation}
\begin{eqnarray}\label{7}
\frac{1}{x}F^{ep}_{2,NLO}(x,Q^{2})=\sum_{q}e_{q}^{2}(q(\xi,Q^{2})+\overline{q}(\xi,Q^{2})
 \nonumber\\+\frac{\alpha_{s}(Q^{2})}{2\pi}[C_{q,2}\otimes(q+\overline{q})+2C_{g,2}\otimes(g)])\;.
\end{eqnarray}
In above equations, $q$ and $ \overline{q} $ refer to quark and
anti-quark distributions with different flavours and electric charge, $ e_{q} $. The
gluon distribution is representing by $ g(x,Q^{2}) $. Wilson
coefficients, $ C_{q,2}(z)$, $ C_{g,2}(z) $ and also the
convolution integral $ C \otimes q $ are given by \cite{10}:
\begin{equation}\label{8}
 C_{q,2}(z)=\frac{4}{3}[\frac{1+z^{2}}{1-z}(ln\frac{1-z}{z}-\frac{3}{4})+\frac{1}{4}(9+5z)]_{+}\;,
\end{equation}
\begin{eqnarray}\label{9}
 C_{g,2}(z)=\frac{1}{2}[(z^{2}+(1-z)^{2})ln\frac{1-z}{z}\;,
 \nonumber\\-1+8z(1-z)]_{+}
\end{eqnarray}
\begin{equation}\label{10}
 C \otimes q=\int^{1}_{x}\frac{dy}{y}C[\frac{x}{y}]q(y,Q^{2})\;.
\end{equation}
The plus sign in Eqs.(\ref{8},\ref{9}) would be appeared in the convolution
integral as in following:
\begin{eqnarray}\label{11}
\int^{1}_{x}\frac{dy}{y}f(\frac{x}{y})_{+}g(y)
\nonumber\\=\int_{x}^{1}\frac{dy}{y}f(\frac{x}{y})[g(y)-\frac{x}{y}g(x)]
\nonumber\\-g(x)\int_{0}^{x} dy f(y)\;.
\end{eqnarray}
Now we can use the re-scaled  parton densities to make the
improved SF at different approximations. In this case we do not
need to use Eq.(\ref{2}) to employ the TMC effect. All the
required considerations  to take into account the mass effects are
done by the re-scaled parton densities.
\section{Discussions and results}
Here we try to use the re-scaled  parton densities instead of
Eq.(\ref{2}) to get the mass effects for the SF. First we do
employ on parton densities the mass correction, using the three
type re-scaling and then we obtain SF at LO and NLO
approximations. We plot in Fig.1 and 2 the results for SF in NLO approximation
at $ Q^{2}=2.7, 3.5, 4.5$ and $ 15\; GeV^{2} $. In
these figures we use three re-scaled prescriptions as we dealt with
them in section 4. We have added as well in these figures the plots for
nucleon SF, $ F_{2} $, resulted from Eq.(\ref{2}) at the
NLO approximations. The results which are obtained from the third
type of re-scaling  are in good agreements with GP approach.
\begin{figure*}[htp]
\begin{center}
\begin{tabular}{cc}
\vspace{-1.5 cm}
\hspace{-1.5 cm}
      {\includegraphics[width=90mm,height=85mm]{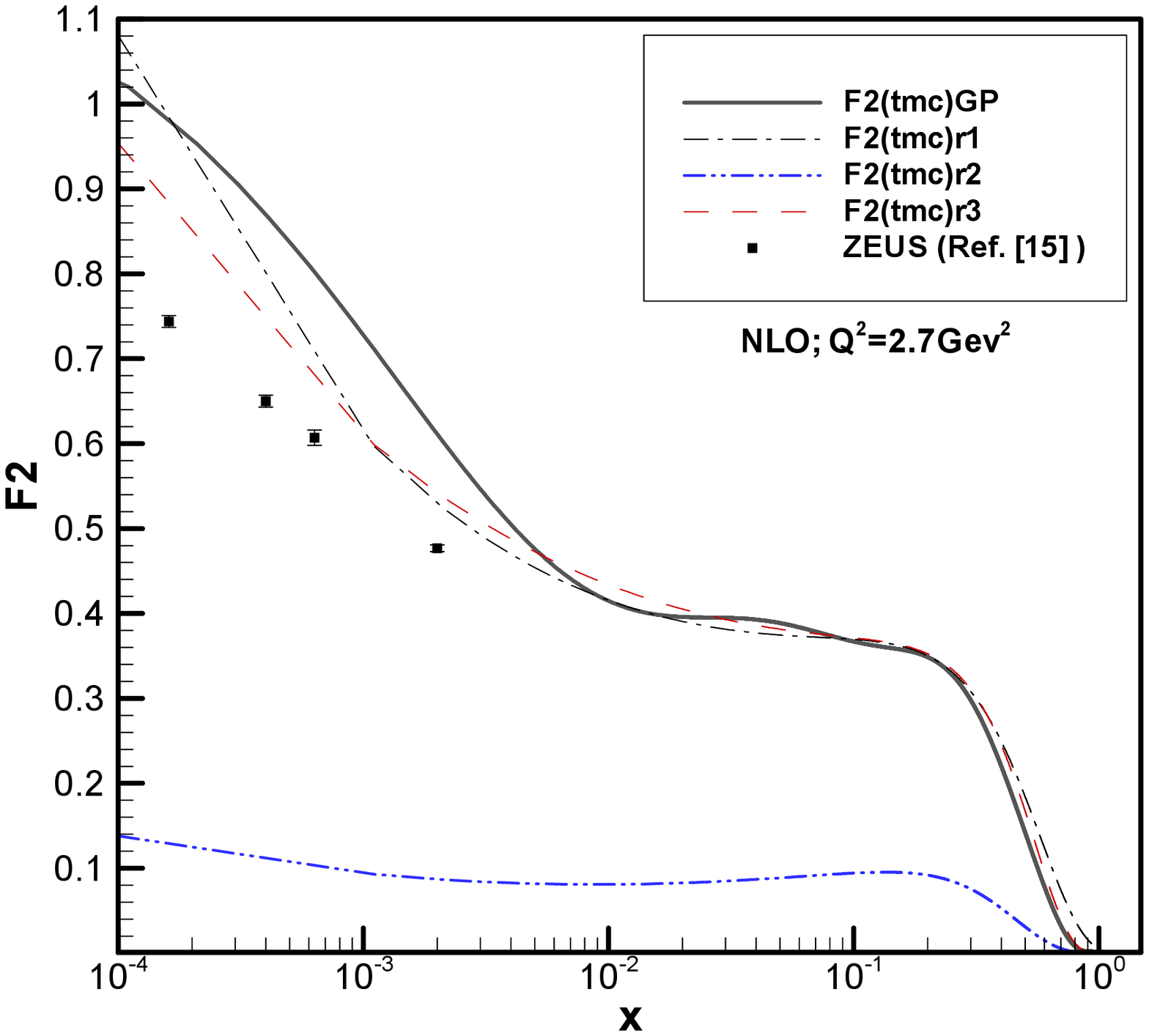}} &
       {\includegraphics[width=90mm,height=85mm]{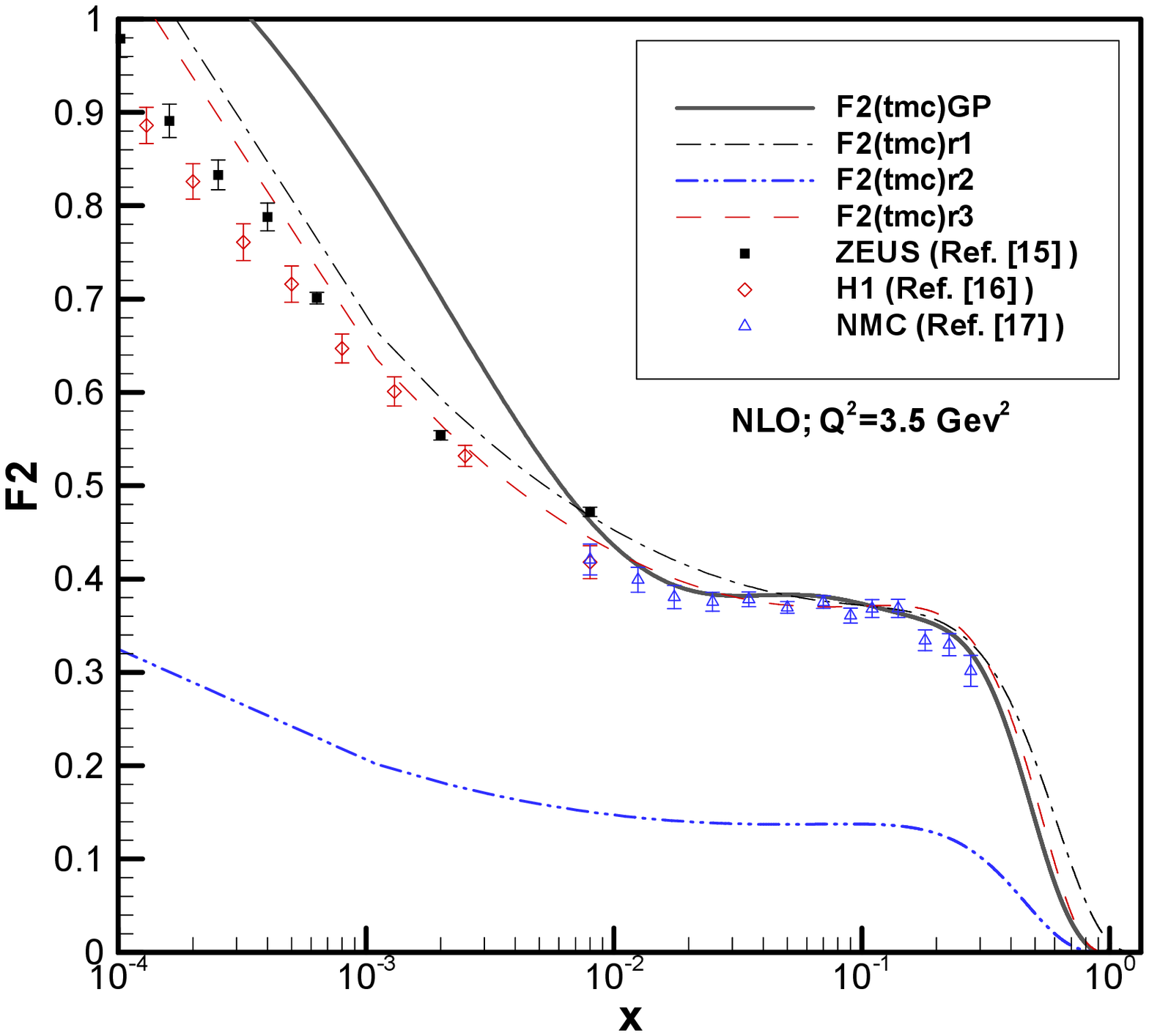}} \\
     \end{tabular}
\caption{ The nucleon structure function at NLO
approximations, resulted from GP  approach \cite{2} and the
re-scaled parton densities at $Q^2=2.7$ and $3.5$ $GeV^2$.
In the legends of these figures $r1$, $r2$ and $r3$ refers to the
first, second and third type of re-scaled parton densities. A comparison with the
available experimental data has also been done
\cite{11,12,13,14,15}.\label{fig12}}
      \end{center}
\end{figure*}
\begin{figure*}[htp]
\begin{center}
\begin{tabular}{cc}
\vspace{-1.5 cm}
\hspace{-1.5 cm}
      {\includegraphics[width=90mm,height=85mm]{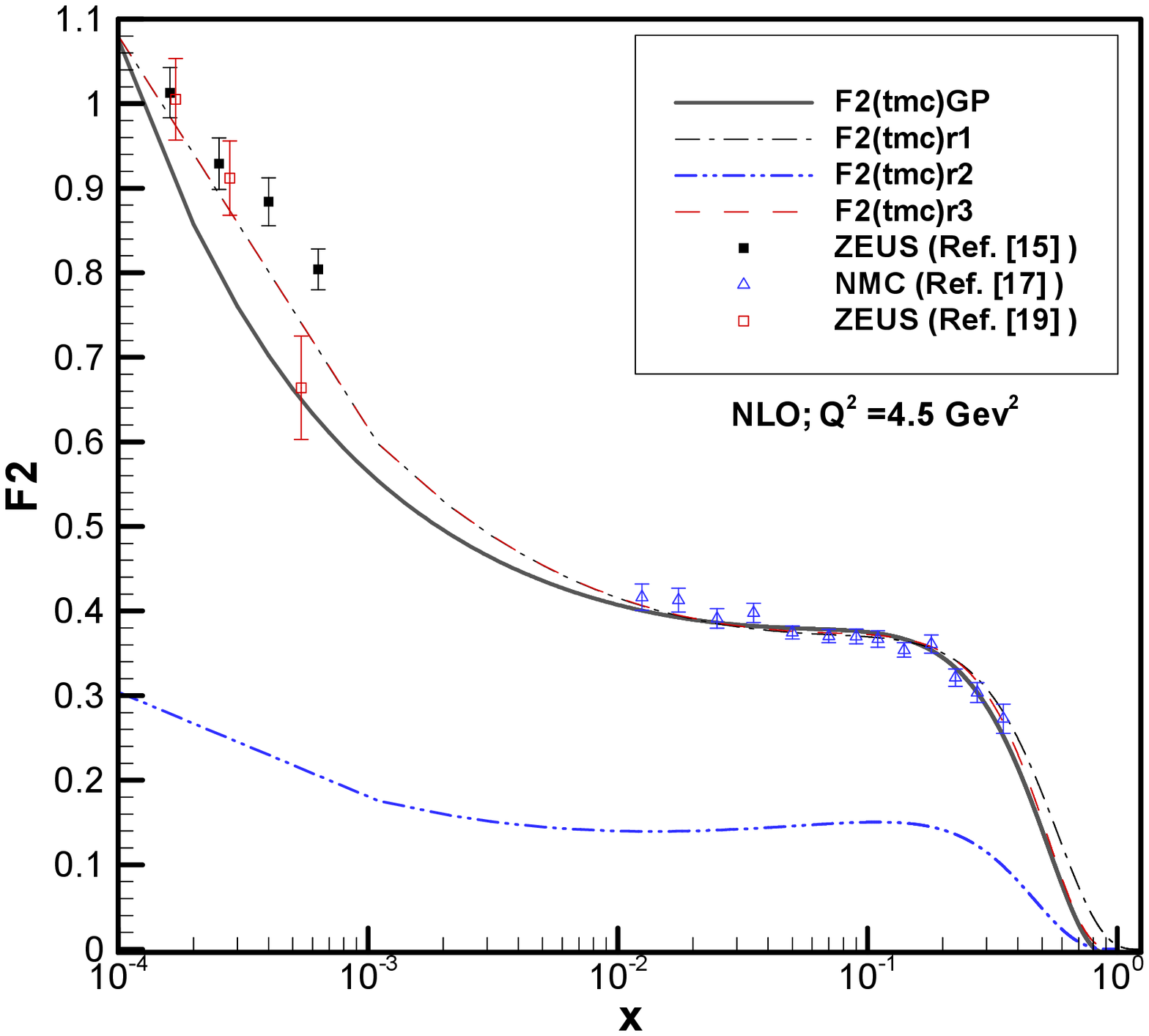}} &
       {\includegraphics[width=90mm,height=85mm]{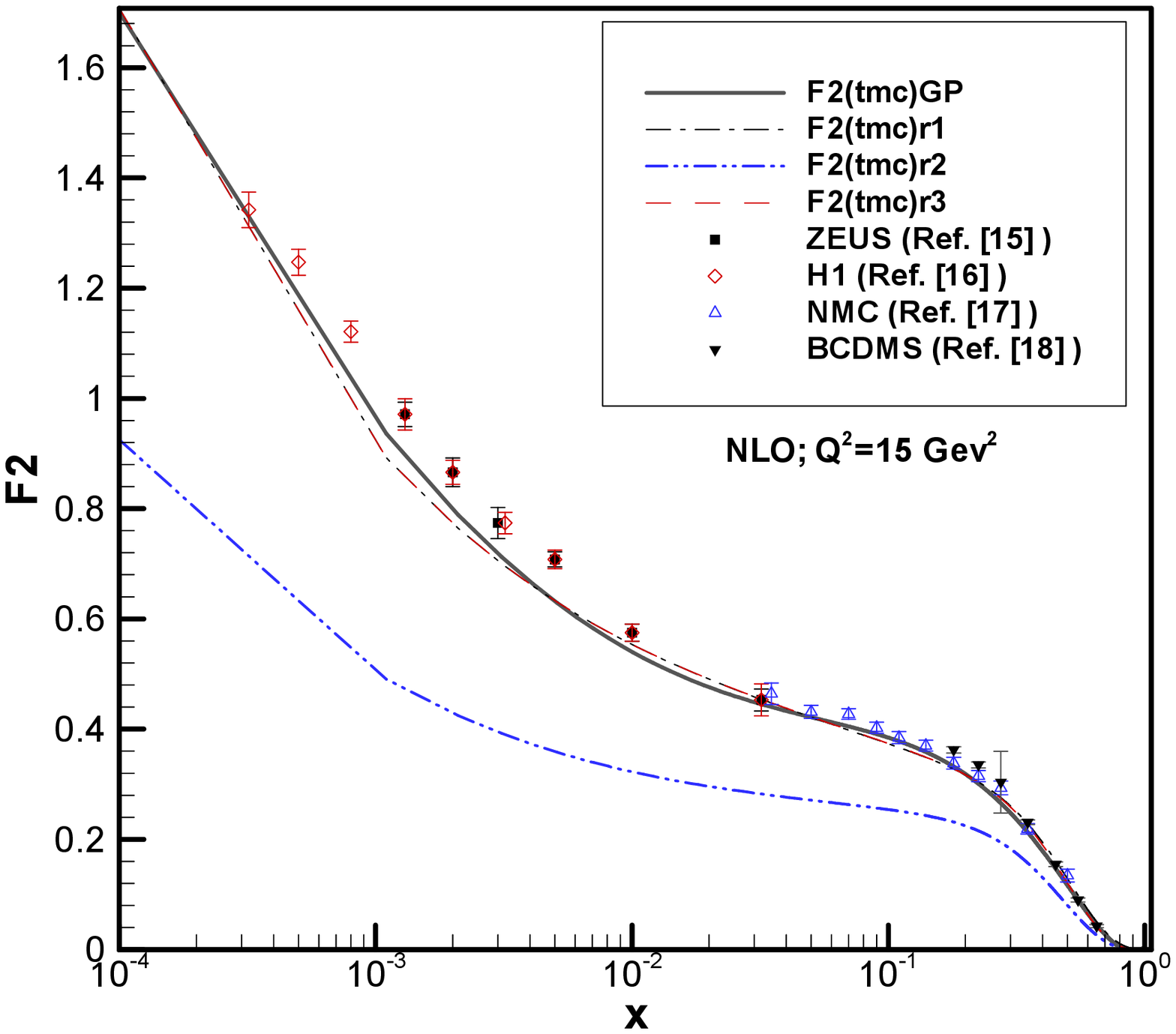}} \\
     \end{tabular}
\caption{As Fig.1 but  at $Q^2=4.5$ and $15$ $GeV^2$. }
      \end{center}
\end{figure*}
A comparison with the available experimental data has also been
done \cite{11,12,13,14,15}. As can be seen, what we get from the
third type of re-scaled parton densities for SF is in better agreement
with GP approach and also the available experimental data. To
show how much the results from GP approach and the third type of re-scaled parton densities
 are near to each other, we plot
in Fig.3 the ratio of structure function using the GP approach and
the third re-scaling method for parton densities which is the best
re-scaling.

However based on the first type re-scaling, there is threshold problem  in Fig.1 and 2 but in order to indicate this problem more
clearly we depict in Fig.4 the nucleon structure function in LO and  NLO
approximations  at typical energy scale $Q^2$=4.5 $GeV^2$
by considering their behavior at large values of $x$ variable. So
in this figure we choose the interval of $x$-Bjorken variable, for instance,
from $0.8$ to maximum value which it can get according to the
relation between Nachtmann variable, $\xi$, and $x$-variable in
Eq.(\ref{1}). We mean that we put the maximum value of $\xi$ which
is 1 in Eq.(\ref{1}) and then depend on the value of $Q^2$, the
maximum value of $x$-variable is obtained respectively. This maximum value of $x$ is
used in depicting the plots of Fig.4.

As can be seen the second and third type of re-scaling does not
contain the threshold problem. The reason for the third re-scaling
is obvious as we explained in Sec.4. For the second type of
re-scaling the $(\xi_0(x=1)-\xi)$ term forbids the
$x$-variable to achieve  values greater than 1. In our suggested third
scaling, again the unchanged $(1-x)$ term in parton
distribution function would prevent  the $x$-variable to get
values grater than 1. Therefore only  for the first type of  re-scaling,  the
$x$-variable can get the amounts greater than 1 by
considering the maximum value of $\xi$ in Eq.(\ref{1}) in
correspond to what we described in above.
\begin{figure*}[htp]
\begin{center}
    \begin{tabular}{cc}
\vspace{-1.5 cm}
\hspace{-1.5 cm}
      {\includegraphics[width=90mm,height=85mm]{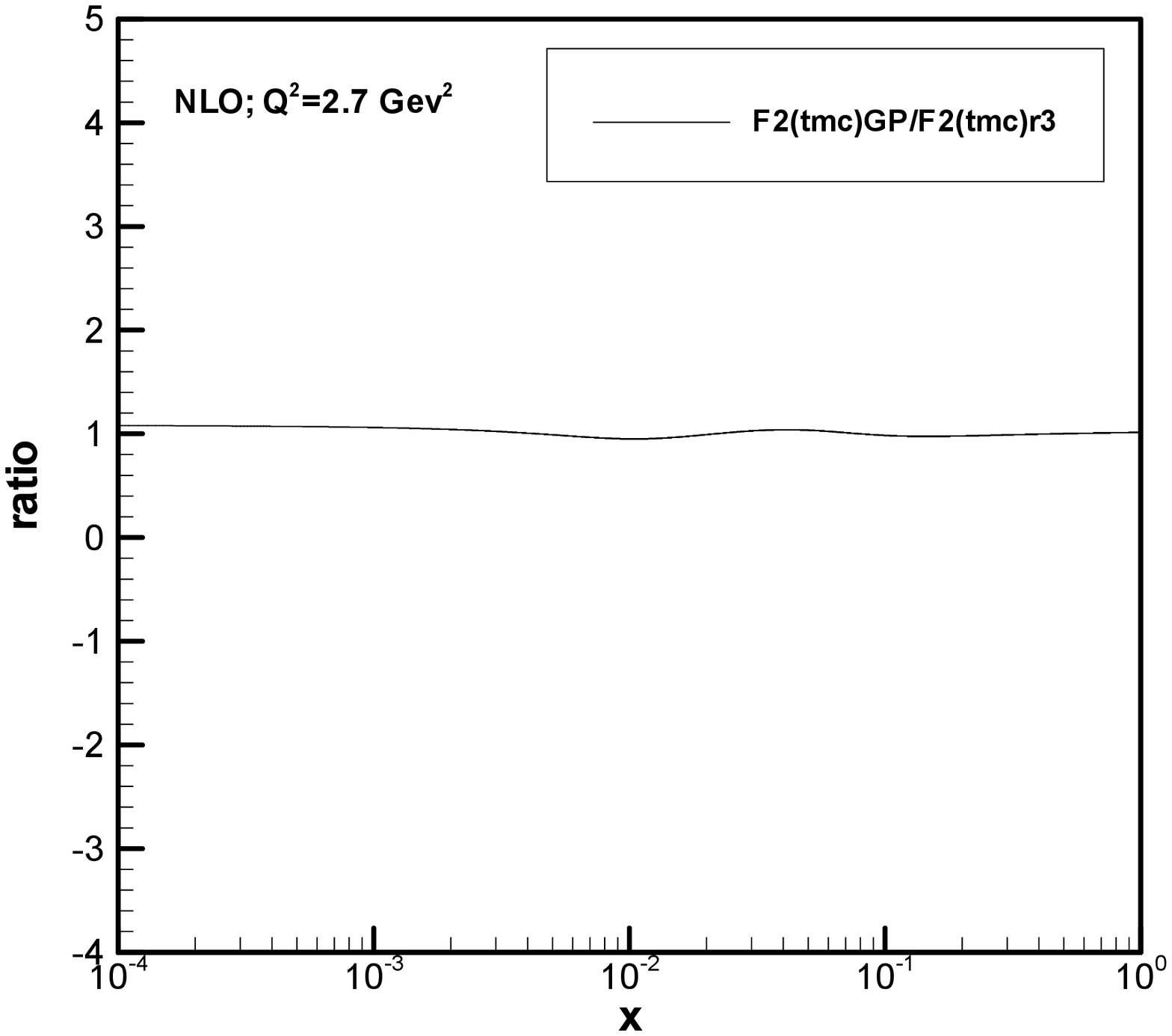}} &
       {\includegraphics[width=90mm,height=85mm]{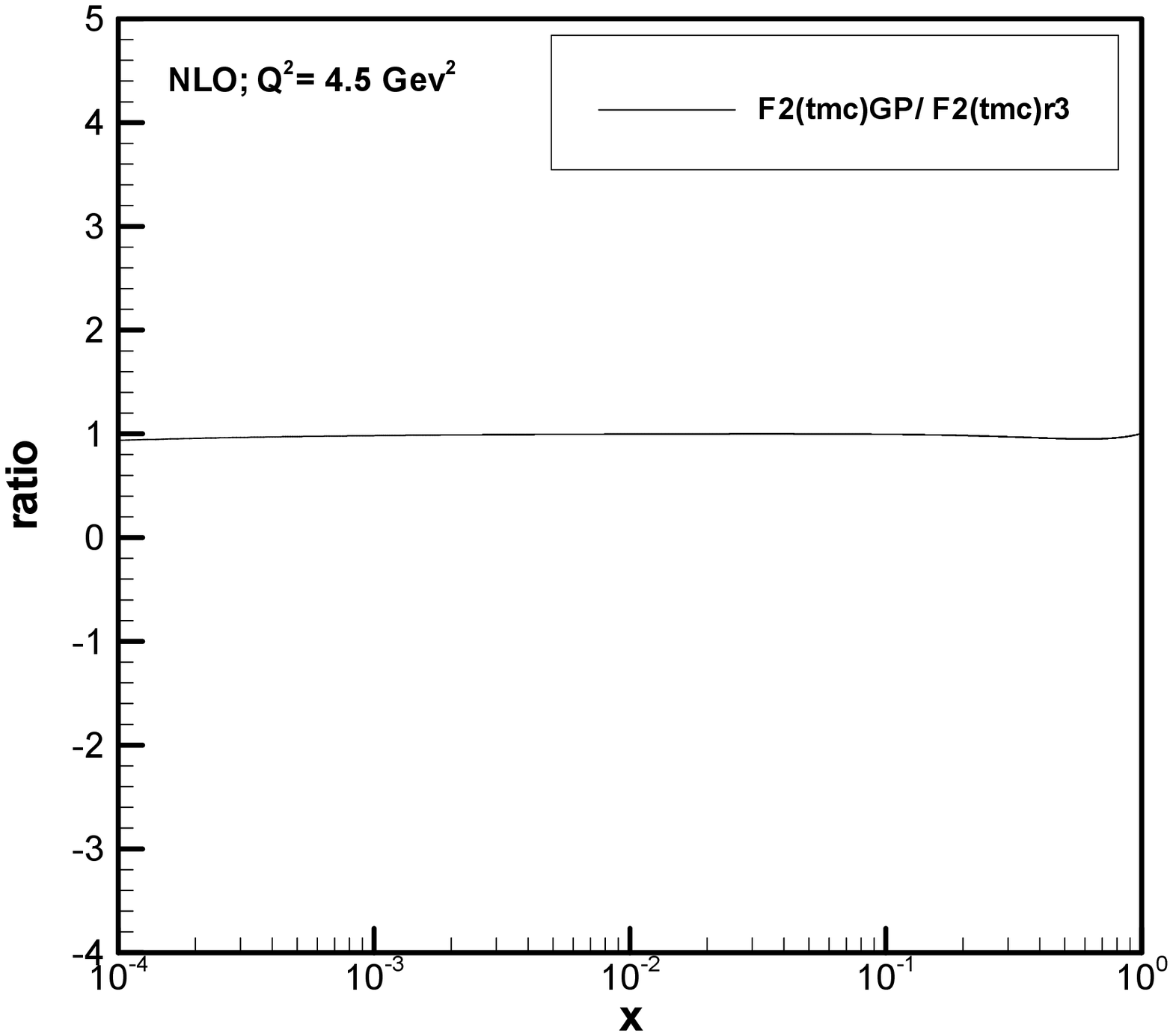}} \\
     \end{tabular}
\caption{The ratio of the nucleon structure function at NLO
approximation, resulted from GP  approach \cite{2} and the
third re-scaled parton densities at
$Q^2=2.7$ and $4.5$ $GeV^2$.}
      \end{center}
\end{figure*}
As referred in above, two solutions are suggested to discard the
threshold problem. The first solution which is given by the second
type re-scaling, contains the $(\xi_0(x=1)-\xi)$ term in parton densities and as
can be seen in Fig.1 and  2 and more clearly in Fig.4, involves the
results which are not compatible with the GP approach and the
first and third re-scalings. The reason is that when in this type
type of re-scaling,  we change the $x$-variable to $\xi$ and in
following do not allow it to get its maximum value ($\xi=1$) but
to get only the maximum value $\xi$=$\xi_0(x=1)$ which obviously is less
than 1, we should be precise to the correlation which exist
between the mass term, $M$, and energy scale, $Q^2$ according to Eq.(\ref{1}). By replacing the $x$ variable  by
$\xi$, we let $\xi$ take its maximum value, 1, when the mass
effect has dominant effect. On the other hand when we have
$\xi$=$\xi_0(x=1)$ as in Eq.(\ref{p3}), it means that we prevent
the mass effect are appeared, since as discussed in previous
section the condition $x=1$ implies that we just encounter with valence
quarks and there are not any other partons like  sea quarks  and gluon.
On the other hand, there is not any interaction in this limit and
therefore we do not expect to have in this case any mass effect.
So the expression $(\xi_0(x=1)-\xi)$ denotes to two conditions which
in fact contradicts each other according to what we explained in
above. This is why in spite that this re-scaling does not involve
threshold problem but its behavior is not compatible with other
scaling and GP approach. In conclusion we can say that this type of
re-scaling (second type) is not a proper one. The first re-scaling
should also be discarded since it contains the threshold problem.
So the only acceptable re-scaling is our suggested scaling which
was called third re-scaling.
\begin{figure*}[htp]
\begin{center}
    \begin{tabular}{cc}
\vspace{-1.5 cm}
\hspace{-1.5 cm}
      {\includegraphics[width=90mm,height=85mm]{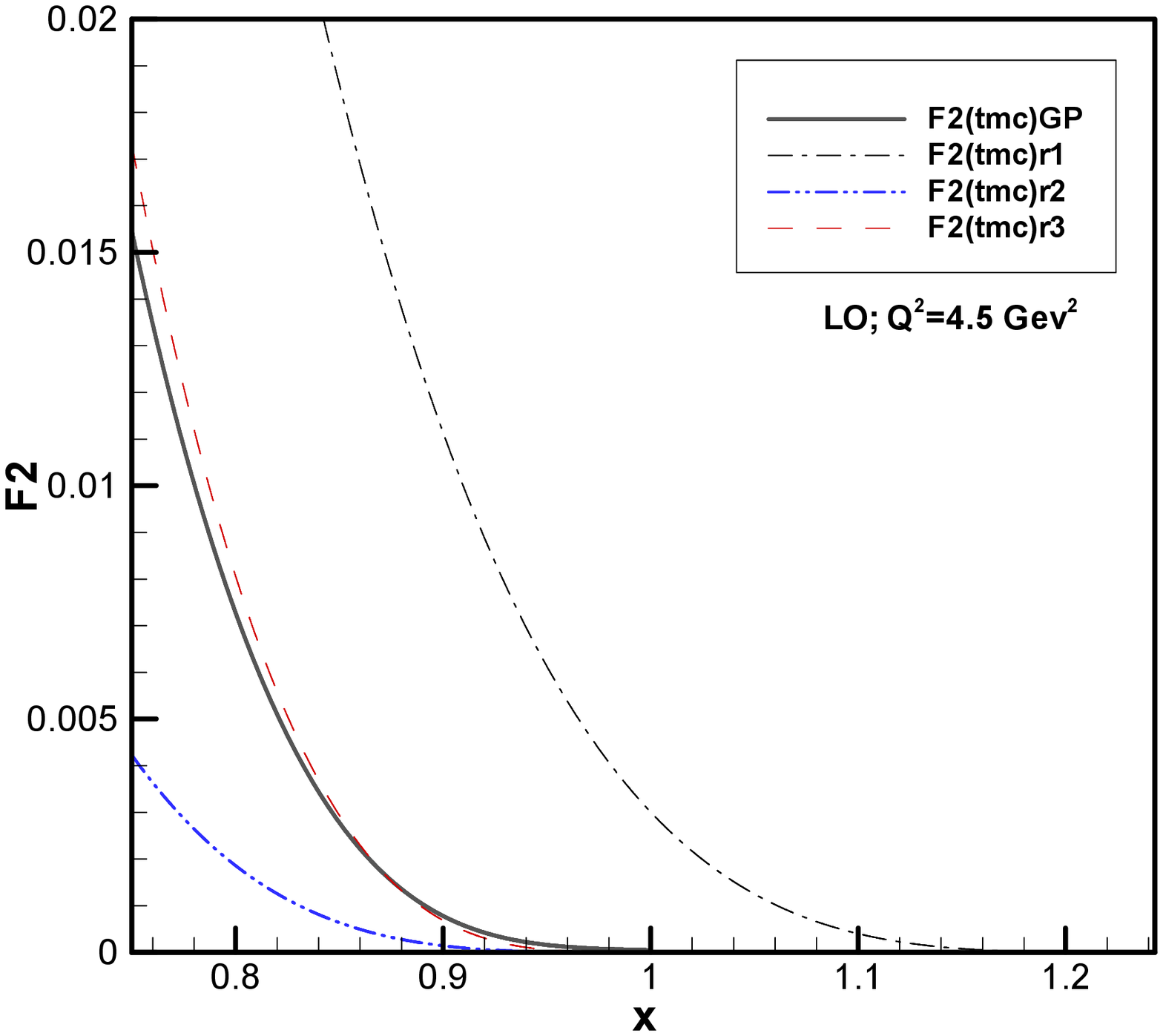}} &
       {\includegraphics[width=90mm,height=85mm]{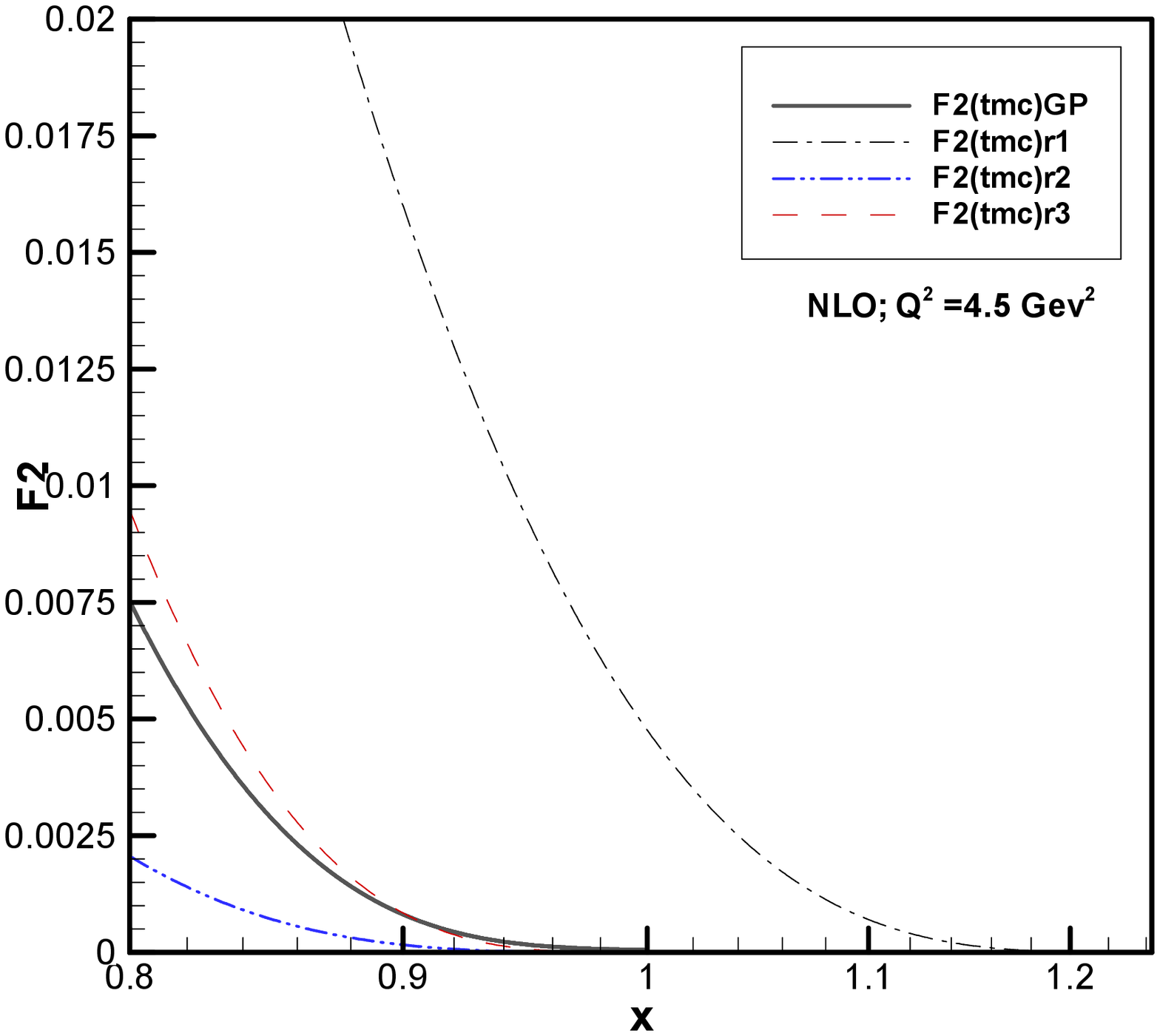}} \\
     \end{tabular}
\caption{The LO and NLO results for the  nucleon structure
function at $Q^2$=4.5 $GeV^2$, resulted from GP
approach \cite{2} and the  re-scaled parton densities for the interval of
0.8$<$x$<$ $x_{max}(\xi=1)$.}
      \end{center}
\end{figure*}

To confirm the validity of discussion in connection to the
deficiency of the second type re-scaling, we can refer to the
Ref.\cite{4} where by looking at Fig.1 and Fig.2 of this
reference, it can be seen that the behavior of moment ratio of
nucleon structure function, resulting from the second type of
re-scaling which was denoted  there as ``Threshold dependent (TD)"
is different with respect to the other used re-scalings in that
paper. Looking  as well at Fig.3 of this reference which is depicting
the structure function, based on different re-scalings, once again
confirms what is expecting  from the second type re-scaling in which
indicates different behavior for SF with respect to the other used
re-scalings in that paper . Fig.5 of this reference, which is
relating to longitudinal structure function, once again indicates
a different behavior of TD re-scaling (second type of re-scaling
in our paper) and therefore can be used as an another evidence to
approve our reasons why this type of re-scaling presents such
different behavior. Consequently we can now claim that the best
candidate for re-scaled parton densities  is the third type  re-scaling which we
suggested in this paper.
\section{Conclusions}
In this paper we investigated a new method to do the mass correction for
the nucleon structure function, rather than to  use the  Eq.(\ref{2})
which contains cumbersome calculations. In this case
the mass correction has been employed directly by parton densities rather than the SF as was done
by the Georgi and Politzer approach \cite{2}.
Three prescriptions for the re-scaled parton
densities were introduced. Using the third type of the re-scaled
parton densities, the threshold problem  has  been removed
properly, so as the SF would be existed only in the physical
region of $x$-Bjorken variable. As we can see from the related figures, what we got  for
the corrected SF considering the TMC effect, using the third type
re-scaling, are very similar to the result of GP approach and even
better than the results when we employ  the other  types of the
re-scaling. In fact, first type re-scaling was discarded since
contains the threshold problem. The second type re-scaling
indicated un-compatible results with respect to the other
re-scaling types. The best candidate for re-scaling IS the
third type re-scaling which was suggested and we described it
extensively in the article.

As a further research task, the TMC effect can be extended  to
the polarized case and then  to the nuclear matter which we hope to give a
report on them in future.

\end{document}